\documentclass[twocolumn,showpacs,preprintnumbers,aps,pra,floatfix,10pt,superscriptaddress]{revtex4-1}
\pdfoutput=1
\usepackage{subfigure}
\usepackage{graphicx}
\usepackage{color}
\usepackage{epsfig}
\usepackage{amsmath}
\usepackage{amssymb}
\usepackage{verbatim}
\usepackage{graphics}
\usepackage{natbib}

\usepackage{float}
\floatstyle{ruled}

\usepackage{pstricks}
\usepackage{eepic}
\def\<{\langle}
\def\>{\rangle}
\def\sd{\Delta}
\def\elle{w}
\begin{document}

\title{Emergence of linear elasticity from the atomistic description of matter}

\author{Abdullah Cakir} 
 \email{acakir@ntu.edu.sg}
\affiliation{Division of Physics and Applied Physics, School of Physical and
  Mathematical Sciences, Nanyang Technological University, Singapore}

\author{Massimo Pica Ciamarra}

\affiliation{Division of Physics and Applied Physics, School of Physical and
    Mathematical Sciences, Nanyang Technological University, Singapore and CNR--SPIN, Dipartimento
    di Scienze Fisiche, Universit\`a di Napoli Federico II, I-80126, Napoli, Italy}

\begin{abstract}
We investigate the emergence of the
  continuum elastic limit from the atomistic description of matter at zero temperature considering
how locally defined elastic quantities depend on the coarse graining length scale.
Results obtained numerically investigating different model systems are rationalized in a unifying
picture according to which the continuum elastic limit emerges through a process
determined by two system properties, the degree of disorder, and a length scale associated
to the transverse low-frequency vibrational modes. The degree of disorder controls the emergence
of long-range local shear stress and shear strain correlations, while the transverse length scale
influences the amplitude of the fluctuations of the local elastic constants.
\end{abstract}
%\pacs
%{
%62.20.-x % Mechanical properties of solids
%81.70.Bt % Mechanical testing, impact tests, static and dynamic loads
%83.80.Fg % Granular solids
%}
\maketitle
\section{Introduction}
The mechanical properties of a solid can be described by taking into account the discrete nature of matter,
via the introduction of an equivalent network of masses and springs.
If the size of the solid is much larger than the typical interatomic spacing,
matter can be treated as a continuum and the mechanical properties are conveniently described
through elasticity theory~\cite{landau1986theory}.
%discrete description can be replaced elastic properties are 
%however 
%Conversely, at larger scales they are described through elasticity theory,~\cite{landau1986theory}
%that treats matter as a continuum. 
Here we consider how the continuum description sets it as the elastic properties
of a material are investigated on increasingly larger spatial scales. This is a question easily addressed
in ideal crystals, where the two descriptions can be related exploiting the periodicity of the
lattice structure.~\cite{landau1986theory,chaikin_principles_2010} In real crystals, that are made
of many randomly oriented ideal micro crystals, periodicity is lost and the connection between the
two descriptions is not obvious; in this case, one tries to recover the continuum description
considering how the micro crystals rotate in response to external deformations~\cite{margulies_situ_2001,ahluwalia_elastic_2003}.
In amorphous materials the problem is more challenging, as
disorder induces non-affine particle displacements~\cite{didonna2005nonaffine}
in response to macroscopic deformations~\cite{didonna2005nonaffine}.
It is known that this non-affine displacement lowers the elastic constants by influencing the
fluctuation term of the stress tensor~\cite{hoover_adiabatic_1969,barrat_elastic_1988}.
However, the role of disorder on the emergence of the continuum limit is unclear. 
To investigate how disorder influences the emergence of linear elasticity,
one needs to consider the dependence of locally defined elastic properties
on the coarse-graining length scale, in systems with different degree of disorder.
Numerically, the measure of local elastic properties 
can be carried out exploiting the derivation of exact
coarse graining expression for local strain and stress
fields~\cite{goldhirsch2002microscopic,goldenberg2007particle,zhang2010coarse},
or via fluctuation formulas\cite{lutsko_stress_1988,lutsko_generalized_1989}.
In this research direction, previous results on a two-dimensional weakly polydisperse Lennard-Jones (LJ)~\cite{tsamados2009local} system,
as well as on systems of spheres interacting via finite range purely repulsive potentials~\cite{mizuno_spatial_2016},
have found the standard deviation (s.d.) of elastic constants defined on a coarse graining length scale $\elle$
to scale as $\elle^{-\alpha}$,
with $\alpha = d/2$ in both~\cite{tsamados2009local} $d = 2$ and $d = 3$~\cite{Hideyuki} spatial
dimensions. The value $\alpha = d/2$ has been rationalized through the central limit theorem,
assuming local elastic constants to be the average of $\propto \elle^d$ random variables
associated to the coarse graining region. 
However, a local modulus is 
not the sum of random variables, but rather the ratio of two coarse-grained quantities, a local stress change and a local strain change, 
so that the applicability of the central limit theorem is questionable. In addition, the central limit theorem
holds in the presence of short range correlations, while the local shear strain could be long-range
correlated, as observed in colloidal~\cite{chikkadi2011long} and metallic~\cite{shang2014evolution}
glasses, and theoretically predicted in granular systems~\cite{henkes_statistical_2009}. These
considerations suggest that the exponent $\alpha$, and therefore the process by which continuum
elasticity sets in, could not be universal.

Here we show, through the numerical investigation of different model systems with different degree of disorder, 
that the process by which continuum elasticity sets in is not universal. Rather, it depends on 
the local order and on a length scale associated to the transverse acoustic modes. The degree of local order fixes the value of the
exponent $\alpha_G$ governing the asymptotic scaling of the s.d. of the local shear modulus $G$
with the coarse-graining length scale, $\sd G \propto \elle^{-\alpha_G}$. We find $\alpha_G = d/2$
only in the presence of a high degree of local order, as in previously investigated weakly
polydisperse~\cite{tsamados2009local} or monodisperse systems~\cite{mizuno_spatial_2016}. In general,
$\alpha_G$ depends on the degree of local order, and its value is understood from the coarse-graining length dependence
of the s.d. of the  local shear stress and of the local shear strain. The transverse
vibrational modes fix the length scale $\xi_G$ above which $\sd_G$ scale as a power law with $\elle$. 
In finite range purely repulsive particle systems this length
scale diverges in the zero pressure limit, as the jamming transition is approached. 
%The structure of
%the paper is as follows. In Sec.~\ref{localelasd} we introduce local elastic quantities
%and discuss their expected scaling with the coarse-graining length. 
%In Sec.~\ref{sec:numeric} we introduce the different numerical models we have considered,
%and detail the numerical procedure we used to measure the local elastic constant. 
%Results are presented in Sec.~\ref{sec:results}.
%We present our results and conclude with
%Section~\ref{conc}.

\section{Local elastic constants}
\label{localelasd}
\subsection{Definitions of Local Quantities}
\label{localdef}
We use the approach developed by Goldhirsch~\cite{goldhirsch2010stress} to define the local stress
and strain based on coarse grained displacement
fields. Local stress~\cite{goldhirsch2002microscopic} and displacement fields~\cite{zhang2010coarse} 
are defined as
\begin{equation}
  \sigma_{ \alpha \beta} ({\bf{r}}) = \frac{1}{2}\sum_{i \neq j} F_{ij \alpha}
  r_{ij \beta} \int_{0}^{1} ds \phi_W({\bf{r}}-{\bf{r_{\it{i}}}}+s {{\bf{r_{\it{ij}}}}})
\label{stres}
\end{equation} 
and as
\begin{equation}
{\bf u}^{\ell}({\bf r})= \frac{\sum\limits_{i} m_i {\bf u}_i \phi_W({\bf r}-{\bf r}_i)}{\sum\limits_{j} m_j \phi_W({\bf r}-{\bf r}_j)}.
\label{disp}
\end{equation} 
Here ${\bf u}_{i}$ is the displacement of particle $i$ induced by
an external deformation, and $\phi_W({\bf r}) = \frac{1}{\pi W^{2}} \exp[{-{\bf r}^{2}/W^{2}}]$ is a 
2d Gaussian coarse graining function~\cite{goldhirsch2010stress}. 
$F_{ij \alpha}$ and $r_{ij \alpha}$ are the $\alpha$ components of contact forces between interacting particles $i$ and $j$,
$r_{ij}=|{\bf{r}}_{ij}|=|{\bf{r}}_i-{\bf{r}}_j|$ is the interparticle separation, $m_{i}$ is the particle mass.
Given the displacement field, the strain $\epsilon_{\alpha \beta}({\bf r})$ is derived as
\begin{equation}
\epsilon_{\alpha \beta}({\bf r})=\frac{1}{2}\left( \frac{\partial u^{\ell}_{\alpha}({\bf r})}{\partial r_{\beta}}+\frac{\partial u^{\ell}_{\beta}({\bf r})}{\partial r_{\alpha}} \right).
\label{stra}
\end{equation} 

We have used the above expressions to determine the local elastic properties on the points of a square grid of lattice spacing $w$,
fixing the standard deviation of the coarse graining function to $W = w/5$, as in previous studies~\cite{goldenberg2006continuum}.
In the following, we consider $w$ to be our coarse-graining length scale.

%In the following, 
%being larger than $W$ so that measures taken at different
%grid points follow linear continuum elasticity~\cite{goldhirsch2002microscopic}.
%are not influenced by the contacts at larger distances than $w$.
%Therefore, the grid points are acting as centers of subsystems of size $w$.
%The grid points are at the centers of the subsystem to prevent the boundary effects.
%This approach of local elasticity
\subsection{Standard Deviations}
\label{stdd}
It is useful to  recapitulate how the s.d. of different locally defined elastic quantities are expected to scale with the coarse
graining length scale.
%To have an insight, we consider many systems divided in $\sim (\frac{L}{w})^d$, where $L$ is the
%system size, equal subsystems as representatives of our systems as in Section~\ref{localdef}.
%We consider how the s.d., $\sd_X$, of a local quantity $X$  is expected to scale with
%coarse graining length scales. 
Eqs.~\ref{stres} and \ref{stra} clarify that the local strain and the local stress are averages of 
properties of contacts and of particles found within the coarse graining region, whose number scales as $w^d$.
According to the central limit theorem their s.d should scale as $w^{-d/2}$ for
$w$ larger than their characteristic correlation length. 
%$-d/2$ indicates uncorrelated $X$,($Y$) across coarse graining lengths. Incidentally, other values
%show $X,Y$ are correlated. 
Conversely, a  local modulus is the ratio of a stress $X$ and of a strain $Y$, $M = X/Y$.
In general, there is not a closed form for the s.d. of the ratio of two random variables.
However, if the probability distributions of $Y$ is such that $P(Y = 0) \simeq 0$, then
\begin{equation}
  \sd_M^2 = \frac{\<X\>^2}{\<Y\>^2}\left[ \frac{\sd_X^2}{\<X\>^2}+\frac{\sd_Y^2}{\<Y\>^2}-2\frac{\sd_{XY}}{\<X\>\<Y\>} \right]
\label{eq:stdexp}
\end{equation}
where $\<X\>$ and $\<Y\>$ are the average values of $X$ and $Y$, and 
$\sd_{XY}$ their covariance.

\section{Numerical Models}
\label{sec:numeric}
We investigate the emergence of linear elasticity in 2d LJ systems, that model
atomic systems, and in 2d Harmonic particles, that model colloids, foams and granular matter,
via zero temperature molecular dynamics simulations~\cite{Plimpton:1995:FPA:201627.201628}.
We consider two LJ systems. The first one is
a mixture of $N_L$ large and $N-N_L$ small particles, with different
fraction $f_L = N_L/N$ of large particles in the range $0$--$1$. We vary $f_L$
to control the degree of disorder, as the size ratio
of the two components is such that when $f_L \simeq 0$ ($\simeq 1$),
our system is polycrystalline, while for $f_L \simeq 0.5$ it is amorphous.
As second LJ system we consider the weakly polydisperse system studied in Ref~\citenum{tsamados2009local},
to rationalize previous results within our approach.
We explore the role of the length scale associated to transverse vibrational modes
considering Harmonic particles at different values of the pressure, 
as this length scale is known to diverge in the zero pressure limit~\cite{silbert_vibrations_2005}.
We notice that previous studies have considered the dependence of the
macroscopic elastic constants on disorder by tuning the relative fraction of different species~\cite{goodrich_solids_2014}.
However, the role of disorder on the emergence of the continuum limit has not been previously addressed.

\subsection{Lennard Jones systems}
In LJ systems particles interact with the potential
\begin{equation*}
 \begin{tabular}{l}
  $V(r_{ij})=4 \epsilon \left[\left(\frac{d_{ij}}{r_{ij}} \right)^{12}-\left(\frac{d_{ij}}{r_{ij}}\right)^{6} +V_{\rm offset} \right]$ \hspace{0. in} {\text{for}} $r_{ij}< r_c$ \\ \\
  $V(r_{ij})=0$ {\text{for}} $r_{ij} \ge r_c$,
  \end{tabular}
\end{equation*}
where $\epsilon=1$ is the energy parameter, $r_c$ is a cutoff, $V_{\rm offset}$ is such that $V(r_c) = 0$,
$d_{ij}=d_i+d_j$, where $d_i$ is the `diameter' of particle $i$.
We consider a bidisperse and a weakly polydisperse system. 
In bidisperse systems, large and small particles have a diameter ratio $1.4$, and $r_c = 2.5d_{ij}$.
In polydisperse systems, particle diameters are drawn from a uniform distribution with 
average value $\<d\>$ and small s.d. $\sim 0.1 \<d\>$, so that the system is weakly disperse~\cite{tsamados2009local},
and the cutoff $r_c =2^{1/6}d_{ij}$ so that the interaction is purely repulsive.
The systems are prepared at zero temperature minimizing the energy of interaction through 
the conjugate-gradient algorithm, starting from infinite temperature states, and have a pressure
of the order of ten. For each set of parameters, our data are obtained averaging over
$100$ independent configurations with $N = 10^4$ particles.

\subsection{Harmonic particle systems}
In Harmonic systems, particles interact via the potential
\begin{equation*}
V(r_{ij})=\frac{1}{2}k(d_{ij}-r_{ij})^{2} {~\text{for}~~} r_{ij}<d_{ij};~~~V(r_{ij})=0 {~\text{otherwise.}}
\end{equation*}
where $k$ is the spring constant, $d_{ij}=d_i+d_j$ is the average diameter of the interacting particles.
We have considered bidisperse systems, with a size ratio $1.4$. 
We have prepared systems at different values of the pressure $p$, 
which is fixed to a high precision ($\left| dp \right|/p < 10^{-2}$)
using a combination of energy minimization and compression/decompression steps.
The final pressure spans $5$ order of magnitudes, $p=10^{-9}-10^{-2}$.

\subsection{Deformations}
We measure local elastic constants by imposing small macroscopic 
compression or shear deformations to the above systems, respectively
to measure local bulk and shear moduli. To explicitly check that we work in the linear response regime,
we increase the strain through a sequence of small steps, minimizing the energy after each step through the conjugate gradient algorithm.
We do observe the locally defined stress and strain
to be linearly related, and thus estimate the local elastic constants
\begin{equation*}
c_{i j}({\bf r},\elle) = {\partial \sigma_{i}({\bf r},\elle)}/{\partial\varepsilon_{j}({\bf r},\elle)},
\end{equation*}
through a linear best-fit of the stress versus strain relation.
Here $c_{ii}$ is to the local bulk modulus, while $c_{ij}$, with $i \neq j$,
is the local shear modulus, under the assumption of rotational symmetry.

% After each deformations in a set of $6$ deformations, the above systems are relaxed back through
% energy minimizations. The pressure of the systems is fixed to a high precision
% ($|\delta p/p| < 10^{-2}$) where $\delta p$ is the change in the pressure after a deformation is
% applied. We implement the procedure described in Section~\ref{localdef}. The resulting diagonal and
% off-diagonal stress components are subtracted from the diagonal  and off-diagonal stress components
% before applications of deformations to extract diagonal and off-diagonal local stress changes for
% uniform compression and shear deformations respectively.  The diagonal and off-diagonal strain
% changes are obtained from the corresponding configuration states of the systems for uniform
% compression and shear deformations respectively.
% 
% The extracted measurements of the
% local stress and strain changes, $\sigma_{\alpha \beta}$ and $\varepsilon_{\alpha \beta}$ are fit with the
% best fit lines where the slopes are the local bulk and shear moduli implementing the local elastic
% constants definition given as
% \begin{equation*}
% c_{i j}({\bf r},\elle) = {\partial \sigma_{i}({\bf r},\elle)}/{\partial\varepsilon_{j}({\bf r},\elle)}
% \end{equation*}
% where we use the notation $i=\alpha \beta$.
% When $\alpha=\beta$, $c_{ii}=K$, the local bulk modulus and $\alpha \neq \beta$, $c_{ii}=G$ for
% notational simplicity. 

\section{Results}
\label{sec:results}
\subsection{Lennard Jones Systems}
\label{lj}
\begin{figure}[t!]
\centering
\includegraphics[scale=0.33]{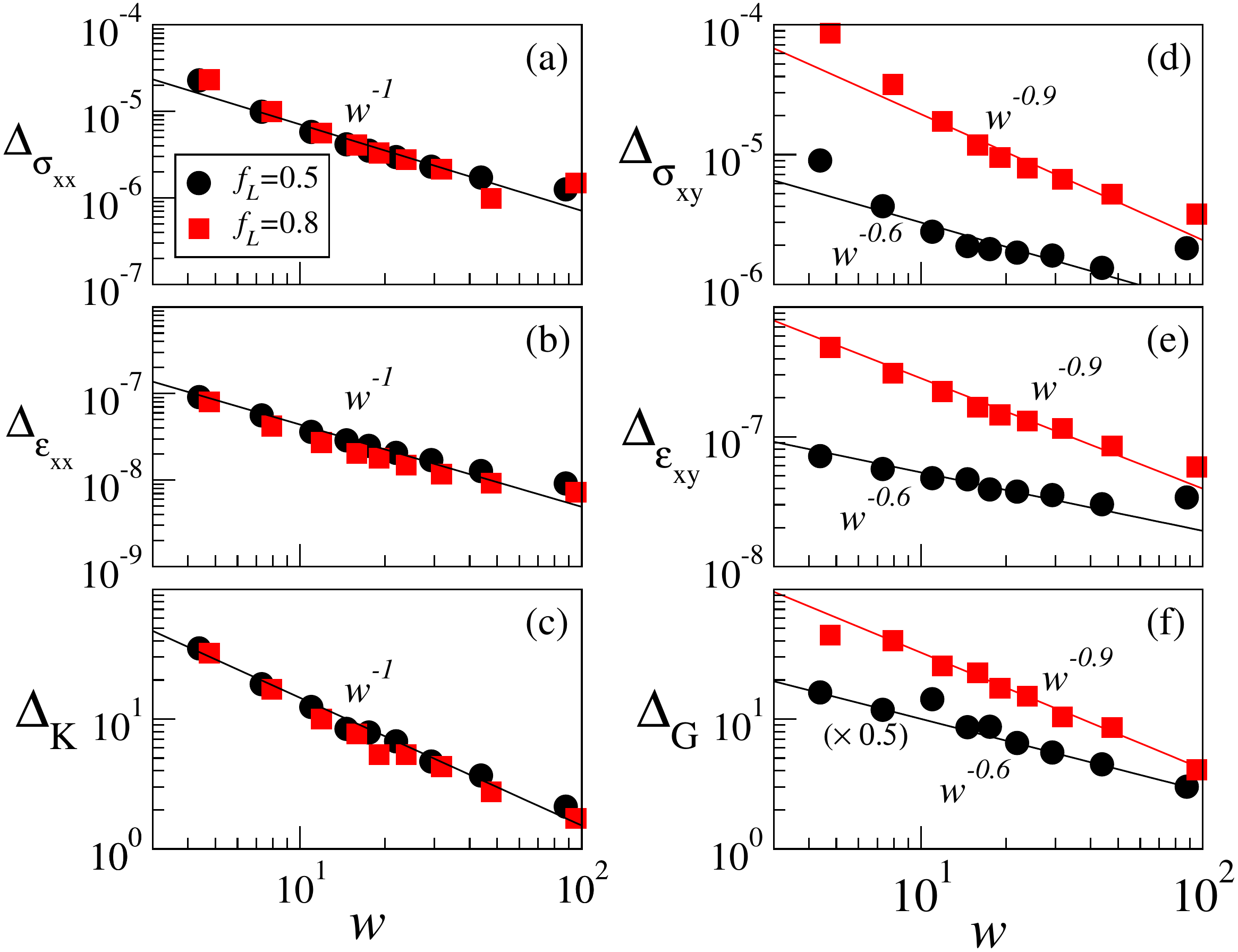}
\caption{\label{fig:std_stress_strain}Dependence of the s.d. of the local stress and of the local
strain, and of the local elastic constants, on the coarse graining length scale.
Panels a,b,c show results for the local diagonal strain, the  local diagonal stress (pressure) and
the local bulk modulus, while panels d,e,f show results
for the local shear strain, the local shear stress and the local shear modulus. 
Data refer to the LJ bidisperse systems with fraction of large particles $f_L$, as indicated.
For $f_L = 0.5$, data in panel $f$ are scaled by $0.5$ for clarity.
}
\end{figure}

The local stress and the local strain are averages of particle
and contact properties in the coarse graining region,
as discussed in Sec.~\ref{stdd}, and therefore their s.d. is expected
to scale as $\elle^{-1}$ if $\elle$ is larger than their correlation length.
In the bidisperse LJ systems, we do actually observe the s.d. of the diagonal local stress, 
$\sigma_{xx}$, and the diagonal local strain, $\varepsilon_{xx}$, to scale as $\elle^{-1}$ regardless of 
the fraction of large particles $f_L$, as illustrated in Fig.~\ref{fig:std_stress_strain}a,b.
Conversely, the s.d. of the local off diagonal (shear) stress,  $\sigma_{xy}$,
and of the local off diagonal (shear) strain, $\varepsilon_{xy}$, scale as $\elle^{-\alpha}$, 
but $\alpha$ depends on $f_L$, as in panels d,e. Specifically,
the exponent $\alpha$ is close to $1$ only at high values of $f_L$.
We also observe $\alpha\simeq 1$ for the weakly polydisperse LJ system, in agreement with earlier
results~\cite{tsamados2009local}.

\begin{figure}[t!]
\centering
\includegraphics[scale=0.5]{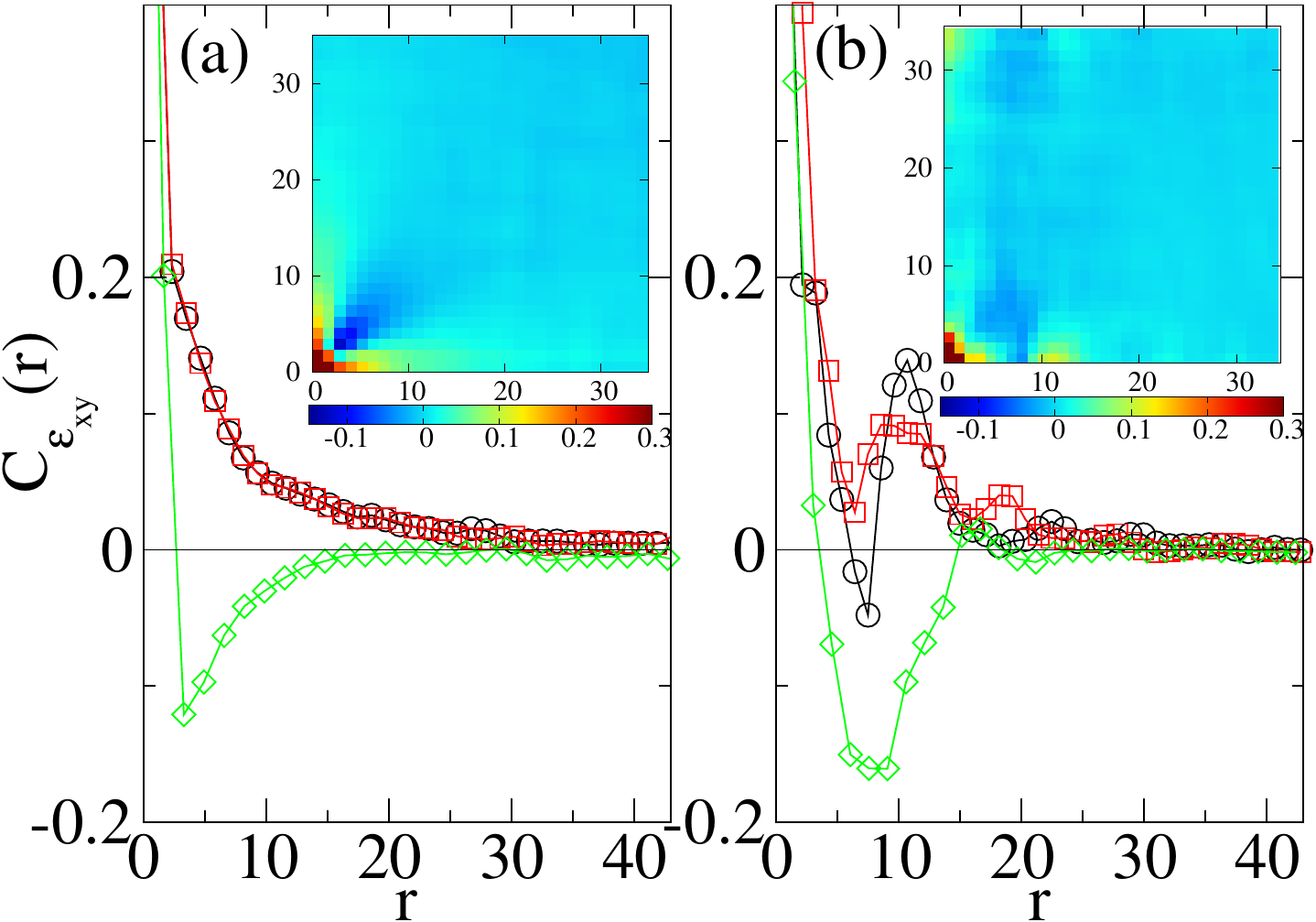} 
\caption{\label{fig:gamma_correlation}Local shear strain correlations of bidisperse LJ systems with $f_L = 0.8$ (a),
and $f_L = 0.5$ (b). The insets show the correlation map in the $xy$ plane. The main
panels illustrate the correlation along the x-axis (circles), the y-axis (squares), and the
diagonal (diamonds). The size of the systems is $L \simeq 80$.
}
\end{figure}

A value $\alpha \neq 1$ implies the breakdown of the central
limit theorem, and hence the presence of long-range correlations in the local shear stress and in
the local shear strain~\cite{chikkadi2011long,shang2014evolution}. 
To explicitly check for the presence of these
long-range correlations we investigate the correlation function of the local shear
strain~\cite{chikkadi2011long} $\varepsilon_{xy}$ defined on the smallest of our investigated
length scale, $\elle = 5$, for two values of $f_L$. Specifically, we have computed the correlation
map
\begin{equation*}
  C_{\varepsilon_{xy}}(x,y) =\frac{\langle \varepsilon_{xy}(0,0)\varepsilon_{xy}(x,y) \rangle - \langle \varepsilon^2_{xy}(0,0) \rangle}{\langle \varepsilon_{xy}(0,0) \rangle^2 - \langle \varepsilon_{xy}^2(0,0) \rangle}
\end{equation*}
and the correlations along the $x$-axis, the $y$-axis, and the diagonal, $C_{\varepsilon_{xy}}(r,0)$, 
$C_{\varepsilon_{xy}}(0,r)$, and $C_{\varepsilon_{xy}}(r/\sqrt{2},r/\sqrt{2})$ respectively, for $f_L = 0.8$ and $0.5$.
For $f_L = 0.8$ we expect no long-range correlations as $\alpha \simeq 1$, and consistently observe the correlation
functions to decay exponentially, as illustrated in Fig.~\ref{fig:gamma_correlation}a.
On the contrary for $f_L = 0.5$ we expect long-range correlations. In this case the correlation
functions exhibit persistent oscillations analogous to those recently experimentally observed in
colloidal systems~\cite{chikkadi2011long}, as illustrated in
Fig.~\ref{fig:gamma_correlation}b.

\begin{figure}[t]
\centering
\includegraphics[scale=0.33]{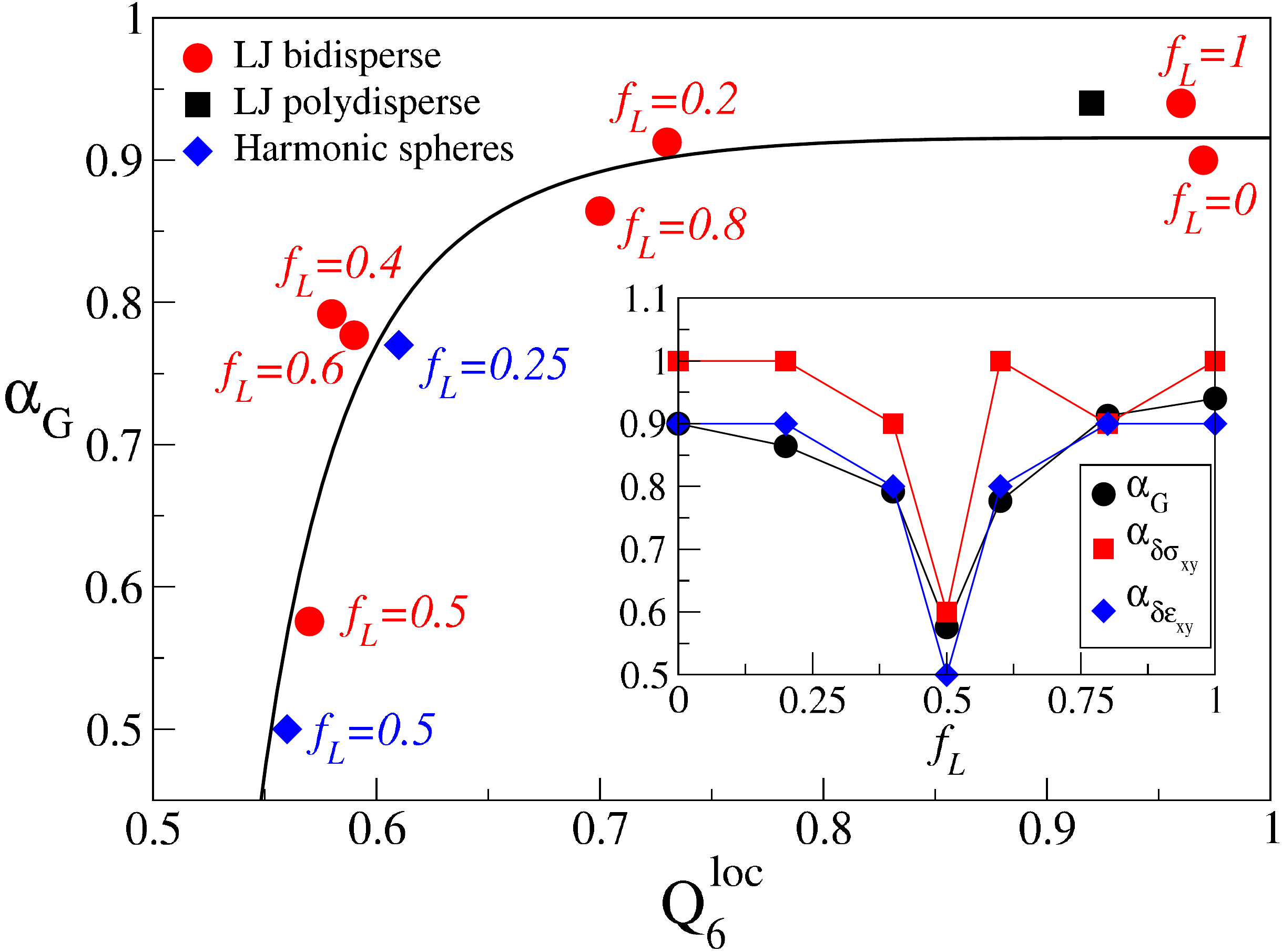}
\caption{\label{fig:ordpcomp}
The inset shows that the fraction $f_L$ of large particle
fixes the exponents governing the scaling of the s.d. of elastic
properties with the coarse-graining length scale $\elle$,
in the bidisperse LJ systems.
Combining results
obtained investigating different systems 
the main figure suggests that the exponent governing the scaling of the s.d. of the local shear
modulus, $\alpha_G$ is fixed by the degree of disorder, as estimated by the local
orientational order parameter $Q_{6}^{loc}$. The panel combines results of 
all three model systems we have have considered. Bidisperse systems are
investigated at different values of the fraction of large particles $f_L$.
}
\end{figure}

Local elastic constants are defined as the ratio of a local stress change and of a local strain
change. Accordingly, as clarified by Eq.~\ref{eq:stdexp},
their s.d. will asymptotically scale as that of the local stress
or as that the local strain, depending on which one asymptotically dominates.
This is consistent with the results of Fig.~\ref{fig:std_stress_strain}c,f,
where we show that the s.d. $\Delta_{\rm K}$ of the local bulk modulus K
scales as $\elle^{-1}$, while that of the local shear modulus scales as 
$\elle^{-\alpha}$, with $\alpha$ depending on $f_L$.
To illustrate the connection between the exponents $\alpha_G$, $\alpha_{\epsilon_{xy}}$,
$\alpha_{\sigma_{xy}}$, with which the s.d. of the local shear modulus, of the local shear strain and
of the local shear stress scale with $w$, we show their dependence on $f_L$ in
Fig.~\ref{fig:ordpcomp}(inset). Since $\alpha_{\epsilon_{xy}} \leq \alpha_{\sigma_{xy}}$ at all
$f_L$, the s.d. of the shear strain dominate that of the shear stress, 
and thus we expect from Eq.~\ref{eq:stdexp} $\alpha_{G} \simeq \alpha_{\epsilon_{xy}}$, as observed.

The inset of Fig.~\ref{fig:ordpcomp} evidences that, in the bidisperse LJ systems,
the exponents depend on the fraction of large particles, $f_L$. This suggests that
the exponents might generally depend on the ordering properties of the system. 
To check whereas this is the case we have investigated
the degree of hexagonal ordering,
which is the ordering expected to arise in 2d, evaluating the local orientational 
order parameter $Q_{{\it m}}^{loc}$
\begin{equation}
Q_{{\it m}}^{loc}=\frac{1}{N}  \sum \limits_{ i=1}^{N_{{\it m}}} \frac{1}{n_{i}} 
  \left| \sum\limits_{j=1}^{n_{i}} e^{ {\it i} {\it m} \theta_{{\bf{r}}_{ij}}} \right|,
 \label{locq} 
\end{equation}
with $m = 6$.
In the above expression, the first sum runs over all the particles, the second one
over all $n_i$ particles in contact with particle $j$, and $\theta_{{\bf{r}}_{ij}}$
is the angle formed by the vector ${\bf{r}}_{ij}$ and a fixed arbitrary one, e.g. the $x$ axis.
We find the exponent $\alpha_G$ corresponding to different systems
to collapse on a same master curve when plotted as a function of the order parameter, as in
Fig.~\ref{fig:ordpcomp}.
Systems with a low degree of disorder, for which $Q_{{\it 6}}^{loc} \simeq 1$, 
have no long range local shear strain or stress correlations, so that the s.d.
of the local shear modulus scales as $\elle^{-\alpha}$ with $\alpha \simeq 1$. As the degree of
disorder increases long range correlation emerges, and $\alpha$ decreases. 
If one assumes linear elasticity to hold to a good approximation above the length scale
$\elle_{\rm LE}$ at which $\Delta_G(\elle_{\rm LE})/\Delta_G(\elle_0) = x$, with $x$ an arbitrary small threshold,
$\elle_0$ a microscopic length scale, then one finds $\elle_{\rm LE} \propto x^{-1/\alpha_G}$. 
Given the dependence of $\alpha_G$ on the disorder
properties of the system, this implies that the length scale $\elle_{\rm LE}$ above which linear elasticity
holds strongly increases as the degree of disorder increases.

\subsection{Disk Particle Systems}
\label{disksres}
Due to the approximate scale--invariance of the interaction potential, in LJ systems pressure plays
the role of a stress scale. Accordingly, the pressure does not influence the ordering properties of the system, and hence
the dependence of the s.d. of the local elastic constants on the coarse graining length scale.
Conversely, in Harmonic systems close to the jamming transition ($p \to 0$) the macroscopic elastic
constants scale critically with the pressure~\cite{o2003jamming}. This makes of interest to
investigate how the jamming transition influences the emergence of the continuum elastic limit. 

The critical-like nature of the jamming transition suggests a scaling relation
for $\sd_G$ and $\sd_K$ of the form,
\begin{equation}
\frac{\sd_G(p,\elle)}{G(p)} = \mathcal G\left(\frac{\elle}{\xi_G(p)}\right); 
\frac{\sd_K(p,\elle)}{K(p)} = \mathcal K\left(\frac{\elle}{\xi_K(p)}\right),
\label{eq:lengtscl}
\end{equation}
where $G \propto p^{1/2}$ and $K \propto p^0$ close to the jamming transition~\cite{o2003jamming}.

\begin{figure}[t]
\centering
\includegraphics[scale=0.33]{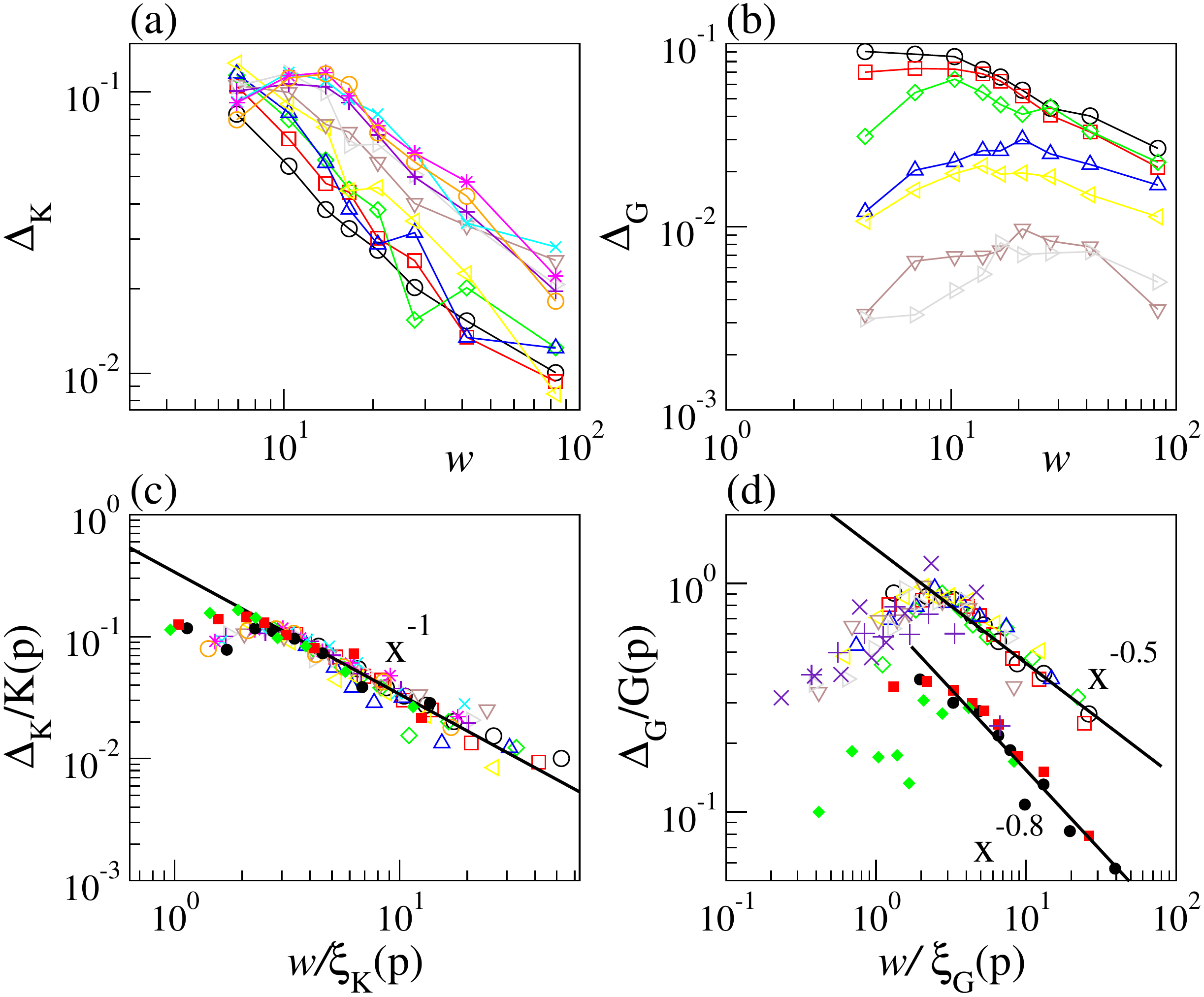}
\caption{\label{fig:stdshearm}Coarse graining length scale dependence of the s.d of the local bulk modulus (a),
for pressures in the range $p= 10^{-2}:1.25\times10^{-7}$, from top to bottom, and of the shear modulus (b), for pressures
in the range $p= 10^{-2}:7\times10^{-5}$, from bottom to top. 
%For smaller pressures shear modulus data areweaffected by finite size effects.
Panels c and d illustrate the collapse of these data according to Eq.~\ref{eq:lengtscl}, with $\xi_K \simeq const.$,
and $\xi_G \propto p^{-0.25}$. All data refer to $f_L = 0.5$ system. In panel (d) we also
illustrate data for $f_L = 0.25$ (full symbols, data scaled by $0.5$), 
to clarify that the length scale $\xi_G$ does not
depend on $f_L$.}

\end{figure}

\begin{figure}[htbp]
  \centering
  \includegraphics[scale=0.225]{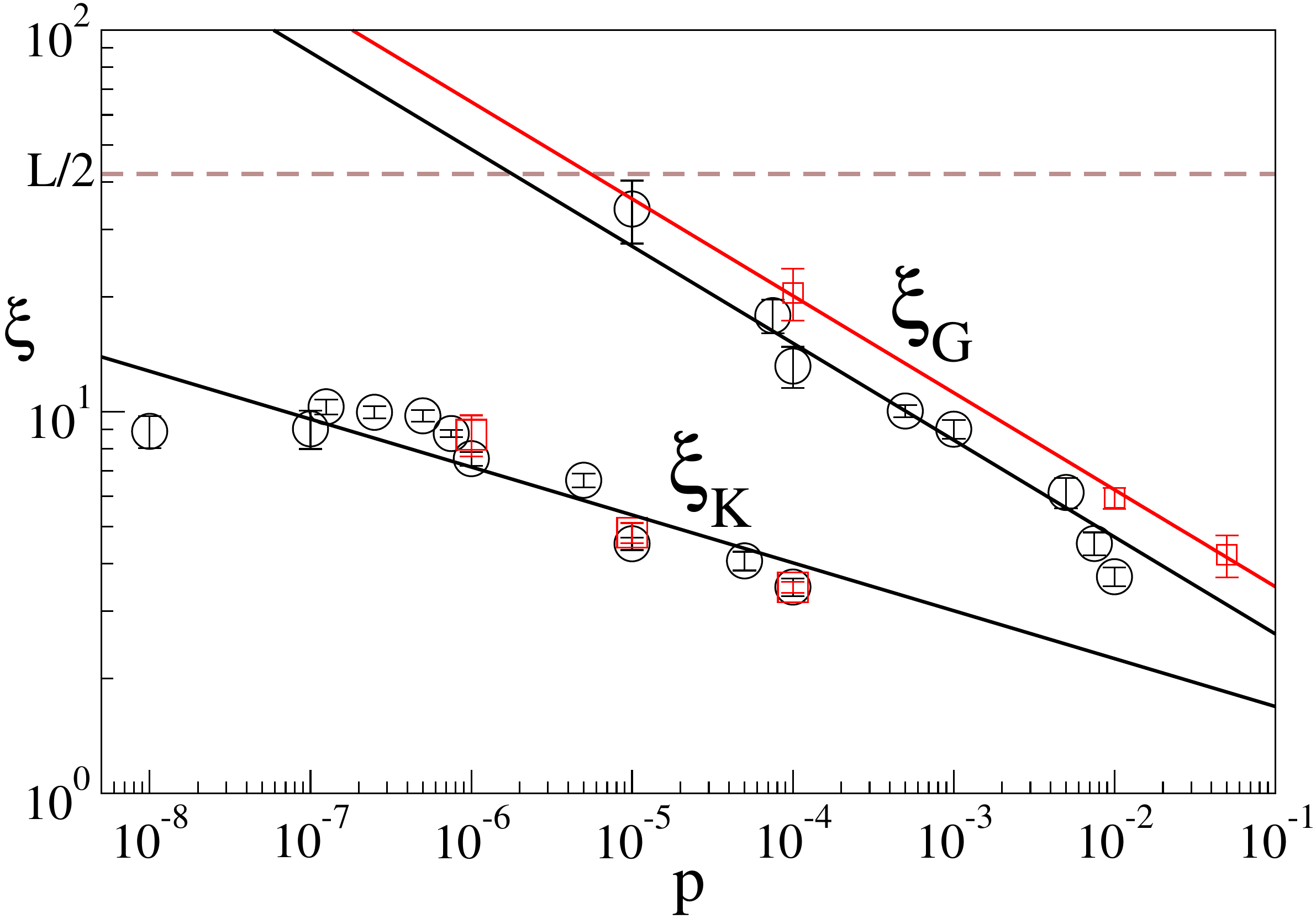}
  \caption{Pressure dependence of the length scales $\xi_{K,G}$ for Harmonic system with different
    $f_{L}$. The solid lines represent power laws, $\nu_{G}=-0.25$.
    Squares (circles) show systems
    with $f_L = 0.25(0.5)$. We do not show data for $\xi_{G}$ at smaller pressures, as they
    are affected by system size effects.
}
  \label{lngthshr}
\end{figure}

We have verified these scaling relations investigating the dependence of the s.d. on $\elle$
for Harmonic disks, and fraction large particles $f_L = 0.5$ and $f_L = 0.25$.
Results for the bulk modulus are shown in Fig.~\ref{fig:stdshearm}a,c.
The s.d. of the local bulk modulus asymptotically scales as $\elle^{-1}$, regardless of the value of $f_L$,
and data corresponding to different pressures can be collapsed as suggested by
the scaling relation of Eq.~\ref{eq:lengtscl}. 
The corresponding length scale $\xi_K$ approaches a constant value in the $p \to 0$ limit, as illustrated in Fig.~\ref{lngthshr}.
Analogous results for the shear modulus are illustrated in Fig.~\ref{fig:stdshearm}b,d.
In this case the asymptotic dependence of the s.d. on $\elle$ depends on $f_L$, as in LJ systems,
and the scaling of data corresponding to different pressures suggests 
$\xi_G \propto p^{-\nu_G}$ where $\nu_G\sim 0.25$, as in Fig.~\ref{lngthshr}.
$\xi_G$ can therefore be identified with the length scale $\xi_c$ associated to the
transverse vibrational modes of jammed systems, that exhibits
the same pressure dependence in systems of Harmonic particles~\cite{silbert_vibrations_2005, lerner2014breakdown}.
Our findings for the scaling of $\Delta_K$ and $\Delta_G$ on $p$, and hence on
the distance from the jamming threshold, 
are in line with recent results~\cite{mizuno_spatial_2016} obtained
determining the local moduli using fluctuation formulas in 
the linear response regime~\cite{lutsko_stress_1988,lutsko_generalized_1989}.

These results clarify that the pressure fixes the length scale above which the
s.d. of the local elastic quantities scales as a power law of the coarse graining
length scale. Conversely, the pressure does not influence 
the asymptotic behavior of the scaling functions, which behave as $\mathcal G(x) \propto x^{-\alpha_G}$,
and as $\mathcal K(x)\propto x^{-1}$, respectively. The investigation of
different values of $f_L$ confirms the influence of the degree of order of the system
on the value of the exponent $\alpha_G$, as summarized in Fig.~\ref{fig:ordpcomp}.

\section{Discussion}
\label{conc}

We have clarified that the process by which the continuum elastic limit emerges,
as evaluated investigating the dependence of the standard deviation of the local
elastic constants on the coarse graining length scale, is not universal.
Rather, two properties of the system at hand influence the dependence of the s.d. of the local
shear modulus with the coarse graining length scale.
Asymptotically, the s.d. scales as a power law of the coarse graining length scale,
with an exponent which is fixed by the degree of disorder.
This is so as the local shear modulus is the ratio of a local shear stress and of 
a local shear strain, that could exhibit long range correlations in the presence of disorder.
The power-law asymptotic regime occurs after a typical length scale, which 
is the typical length scale of the transverse vibrational modes.
The role of this length scale is particularly relevant in systems
of particles interacting via purely repulsive forces, as it diverges in the zero pressure 
limit corresponding to the jamming transition.

The main open question ahead is the individuation of the physical
mechanisms responsible for the reported long range correlations in 
the elastic fields. There are contrasting results in the literature,
as the local elastic moduli have been observed to be Gaussian
in simulations of polymeric glasses~\cite{yoshimoto_mechanical_2004}
and of LJ glasses~\cite{PhysRevE.87.042306,tsamados2009local},
while the local shear strain has been observed to be long range
in colloidal~\cite{chikkadi2011long} and metallic~\cite{shang2014evolution}
glasses, and theoretically predicted in granular systems~\cite{henkes_statistical_2009}. 
Our results suggest that the investigation of the degree of disorder, or
of other structural order parameter such as the excess entropy~\cite{tanaka_critical-like_2010}, could
allow to rationalize this discrepancy. 
%\bibliography{localel}

\begin{thebibliography}{28}%
\makeatletter
\providecommand \@ifxundefined [1]{%
 \@ifx{#1\undefined}
}%
\providecommand \@ifnum [1]{%
 \ifnum #1\expandafter \@firstoftwo
 \else \expandafter \@secondoftwo
 \fi
}%
\providecommand \@ifx [1]{%
 \ifx #1\expandafter \@firstoftwo
 \else \expandafter \@secondoftwo
 \fi
}%
\providecommand \natexlab [1]{#1}%
\providecommand \enquote  [1]{``#1''}%
\providecommand \bibnamefont  [1]{#1}%
\providecommand \bibfnamefont [1]{#1}%
\providecommand \citenamefont [1]{#1}%
\providecommand \href@noop [0]{\@secondoftwo}%
\providecommand \href [0]{\begingroup \@sanitize@url \@href}%
\providecommand \@href[1]{\@@startlink{#1}\@@href}%
\providecommand \@@href[1]{\endgroup#1\@@endlink}%
\providecommand \@sanitize@url [0]{\catcode `\\12\catcode `\$12\catcode
  `\&12\catcode `\#12\catcode `\^12\catcode `\_12\catcode `\%12\relax}%
\providecommand \@@startlink[1]{}%
\providecommand \@@endlink[0]{}%
\providecommand \url  [0]{\begingroup\@sanitize@url \@url }%
\providecommand \@url [1]{\endgroup\@href {#1}{\urlprefix }}%
\providecommand \urlprefix  [0]{URL }%
\providecommand \Eprint [0]{\href }%
\providecommand \doibase [0]{http://dx.doi.org/}%
\providecommand \selectlanguage [0]{\@gobble}%
\providecommand \bibinfo  [0]{\@secondoftwo}%
\providecommand \bibfield  [0]{\@secondoftwo}%
\providecommand \translation [1]{[#1]}%
\providecommand \BibitemOpen [0]{}%
\providecommand \bibitemStop [0]{}%
\providecommand \bibitemNoStop [0]{.\EOS\space}%
\providecommand \EOS [0]{\spacefactor3000\relax}%
\providecommand \BibitemShut  [1]{\csname bibitem#1\endcsname}%
\let\auto@bib@innerbib\@empty
%</preamble>
\bibitem [{\citenamefont {Landau}\ and\ \citenamefont
  {Lifshitz}(1986)}]{landau1986theory}%
  \BibitemOpen
  \bibfield  {author} {\bibinfo {author} {\bibfnamefont {L.~D.}\ \bibnamefont
  {Landau}}\ and\ \bibinfo {author} {\bibfnamefont {E.~M.}\ \bibnamefont
  {Lifshitz}},\ }\href@noop {} {\emph {\bibinfo {title} {Theory of
  Elasticity}}}\ (\bibinfo  {publisher} {Elsevier},\ \bibinfo {address}
  {Oxford},\ \bibinfo {year} {1986})\BibitemShut {NoStop}%
\bibitem [{\citenamefont {Chaikin}\ and\ \citenamefont
  {Lubensky}(2010)}]{chaikin_principles_2010}%
  \BibitemOpen
  \bibfield  {author} {\bibinfo {author} {\bibfnamefont {P.~M.}\ \bibnamefont
  {Chaikin}}\ and\ \bibinfo {author} {\bibfnamefont {T.~C.}\ \bibnamefont
  {Lubensky}},\ }\href@noop {} {\emph {\bibinfo {title}
  {Principles of condensed matter physics}}},\ \bibinfo {edition} {1st}\ ed.\
  (\bibinfo  {publisher} {Cambridge Univ. Press},\ \bibinfo {address}
  {Cambridge},\ \bibinfo {year} {2010})\BibitemShut {NoStop}%
\bibitem [{\citenamefont {Margulies}\ \emph {et~al.}(2001)\citenamefont
  {Margulies}, \citenamefont {Winther},\ and\ \citenamefont
  {Poulsen}}]{margulies_situ_2001}%
  \BibitemOpen
  \bibfield  {author} {\bibinfo {author} {\bibfnamefont {L.}~\bibnamefont
  {Margulies}}, \bibinfo {author} {\bibfnamefont {G.}~\bibnamefont {Winther}},
  \ and\ \bibinfo {author} {\bibfnamefont {H.~F.}\ \bibnamefont {Poulsen}},\
  }\href {\doibase 10.1126/science.1057956} {\bibfield  {journal} {\bibinfo
  {journal} {Science}\ }\textbf {\bibinfo {volume} {291}},\ \bibinfo {pages}
  {2392} (\bibinfo {year} {2001})}\BibitemShut {NoStop}%
\bibitem [{\citenamefont {Ahluwalia}\ \emph {et~al.}(2003)\citenamefont
  {Ahluwalia}, \citenamefont {Lookman},\ and\ \citenamefont
  {Saxena}}]{ahluwalia_elastic_2003}%
  \BibitemOpen
  \bibfield  {author} {\bibinfo {author} {\bibfnamefont {R.}~\bibnamefont
  {Ahluwalia}}, \bibinfo {author} {\bibfnamefont {T.}~\bibnamefont {Lookman}},
  \ and\ \bibinfo {author} {\bibfnamefont {A.}~\bibnamefont {Saxena}},\ }\href
  {\doibase 10.1103/PhysRevLett.91.055501} {\bibfield  {journal} {\bibinfo
  {journal} {Phys. Rev. Lett.}\ }\textbf {\bibinfo {volume} {91}},\ \bibinfo
  {pages} {055501} (\bibinfo {year} {2003})}\BibitemShut {NoStop}%
\bibitem [{\citenamefont {DiDonna}\ and\ \citenamefont
  {Lubensky}(2005)}]{didonna2005nonaffine}%
  \BibitemOpen
  \bibfield  {author} {\bibinfo {author} {\bibfnamefont {B.~A.}\ \bibnamefont
  {DiDonna}}\ and\ \bibinfo {author} {\bibfnamefont {T.~C.}\ \bibnamefont
  {Lubensky}},\ }\href@noop {} {\bibfield  {journal} {\bibinfo  {journal}
  {Physical Review E}\ }\textbf {\bibinfo {volume} {72}},\ \bibinfo {pages}
  {066619} (\bibinfo {year} {2005})}\BibitemShut {NoStop}%
\bibitem [{\citenamefont {Hoover}\ \emph {et~al.}(1969)\citenamefont {Hoover},
  \citenamefont {Holt},\ and\ \citenamefont {Squire}}]{hoover_adiabatic_1969}%
  \BibitemOpen
  \bibfield  {author} {\bibinfo {author} {\bibfnamefont {W.~G.}\ \bibnamefont
  {Hoover}}, \bibinfo {author} {\bibfnamefont {A.~C.}\ \bibnamefont {Holt}}, \
  and\ \bibinfo {author} {\bibfnamefont {D.~R.}\ \bibnamefont {Squire}},\
  }\href {\doibase 10.1016/0031-8914(69)90217-1} {\bibfield  {journal}
  {\bibinfo  {journal} {Physica}\ }\textbf {\bibinfo {volume} {44}},\ \bibinfo
  {pages} {437} (\bibinfo {year} {1969})}\BibitemShut {NoStop}%
\bibitem [{\citenamefont {Barrat}\ \emph {et~al.}(1988)\citenamefont {Barrat},
  \citenamefont {Roux}, \citenamefont {Hansen},\ and\ \citenamefont
  {Klein}}]{barrat_elastic_1988}%
  \BibitemOpen
  \bibfield  {author} {\bibinfo {author} {\bibfnamefont {J.-L.}\ \bibnamefont
  {Barrat}}, \bibinfo {author} {\bibfnamefont {J.-N.}\ \bibnamefont {Roux}},
  \bibinfo {author} {\bibfnamefont {J.-P.}\ \bibnamefont {Hansen}}, \ and\
  \bibinfo {author} {\bibfnamefont {M.~L.}\ \bibnamefont {Klein}},\ }\href
  {\doibase 10.1209/0295-5075/7/8/007} {\bibfield  {journal} {\bibinfo
  {journal} {Europhysics Letters (EPL)}\ }\textbf {\bibinfo {volume} {7}},\
  \bibinfo {pages} {707} (\bibinfo {year} {1988})}\BibitemShut {NoStop}%
\bibitem [{\citenamefont {Goldhirsch}\ and\ \citenamefont
  {Goldenberg}(2002)}]{goldhirsch2002microscopic}%
  \BibitemOpen
  \bibfield  {author} {\bibinfo {author} {\bibfnamefont {I.}~\bibnamefont
  {Goldhirsch}}\ and\ \bibinfo {author} {\bibfnamefont {C.}~\bibnamefont
  {Goldenberg}},\ }\href@noop {} {\bibfield  {journal} {\bibinfo  {journal}
  {The European Physical Journal E}\ }\textbf {\bibinfo {volume} {9}},\
  \bibinfo {pages} {245} (\bibinfo {year} {2002})}\BibitemShut {NoStop}%
\bibitem [{\citenamefont {Goldenberg}\ \emph {et~al.}(2007)\citenamefont
  {Goldenberg}, \citenamefont {Tanguy},\ and\ \citenamefont
  {Barrat}}]{goldenberg2007particle}%
  \BibitemOpen
  \bibfield  {author} {\bibinfo {author} {\bibfnamefont {C.}~\bibnamefont
  {Goldenberg}}, \bibinfo {author} {\bibfnamefont {A.}~\bibnamefont {Tanguy}},
  \ and\ \bibinfo {author} {\bibfnamefont {J.-L.}\ \bibnamefont {Barrat}},\
  }\href@noop {} {\bibfield  {journal} {\bibinfo  {journal} {EPL (Europhysics
  Letters)}\ }\textbf {\bibinfo {volume} {80}},\ \bibinfo {pages} {16003}
  (\bibinfo {year} {2007})}\BibitemShut {NoStop}%
\bibitem [{\citenamefont {Zhang}\ \emph {et~al.}(2010)\citenamefont {Zhang},
  \citenamefont {Behringer},\ and\ \citenamefont
  {Goldhirsch}}]{zhang2010coarse}%
  \BibitemOpen
  \bibfield  {author} {\bibinfo {author} {\bibfnamefont {J.}~\bibnamefont
  {Zhang}}, \bibinfo {author} {\bibfnamefont {R.~P.}\ \bibnamefont
  {Behringer}}, \ and\ \bibinfo {author} {\bibfnamefont {I.}~\bibnamefont
  {Goldhirsch}},\ }\href@noop {} {\bibfield  {journal} {\bibinfo  {journal}
  {Progress of Theoretical Physics Supplement}\ }\textbf {\bibinfo {volume}
  {184}},\ \bibinfo {pages} {16} (\bibinfo {year} {2010})}\BibitemShut
  {NoStop}%
\bibitem [{\citenamefont {Lutsko}(1988)}]{lutsko_stress_1988}%
  \BibitemOpen
  \bibfield  {author} {\bibinfo {author} {\bibfnamefont {J.~F.}\ \bibnamefont
  {Lutsko}},\ }\href {\doibase 10.1063/1.341877} {\bibfield  {journal}
  {\bibinfo  {journal} {Journal of Applied Physics}\ }\textbf {\bibinfo
  {volume} {64}},\ \bibinfo {pages} {1152} (\bibinfo {year}
  {1988})}\BibitemShut {NoStop}%
\bibitem [{\citenamefont {Lutsko}(1989)}]{lutsko_generalized_1989}%
  \BibitemOpen
  \bibfield  {author} {\bibinfo {author} {\bibfnamefont {J.~F.}\ \bibnamefont
  {Lutsko}},\ }\href {\doibase 10.1063/1.342716} {\bibfield  {journal}
  {\bibinfo  {journal} {Journal of Applied Physics}\ }\textbf {\bibinfo
  {volume} {65}},\ \bibinfo {pages} {2991} (\bibinfo {year}
  {1989})}\BibitemShut {NoStop}%
\bibitem [{\citenamefont {Tsamados}\ \emph {et~al.}(2009)\citenamefont
  {Tsamados}, \citenamefont {Tanguy}, \citenamefont {Goldenberg},\ and\
  \citenamefont {Barrat}}]{tsamados2009local}%
  \BibitemOpen
  \bibfield  {author} {\bibinfo {author} {\bibfnamefont {M.}~\bibnamefont
  {Tsamados}}, \bibinfo {author} {\bibfnamefont {A.}~\bibnamefont {Tanguy}},
  \bibinfo {author} {\bibfnamefont {C.}~\bibnamefont {Goldenberg}}, \ and\
  \bibinfo {author} {\bibfnamefont {J.-L.}\ \bibnamefont {Barrat}},\
  }\href@noop {} {\bibfield  {journal} {\bibinfo  {journal} {Physical Review
  E}\ }\textbf {\bibinfo {volume} {80}},\ \bibinfo {pages} {026112} (\bibinfo
  {year} {2009})}\BibitemShut {NoStop}%
\bibitem [{\citenamefont {Mizuno}\ \emph {et~al.}(2016)\citenamefont {Mizuno},
  \citenamefont {Silbert},\ and\ \citenamefont {Sperl}}]{mizuno_spatial_2016}%
  \BibitemOpen
  \bibfield  {author} {\bibinfo {author} {\bibfnamefont {H.}~\bibnamefont
  {Mizuno}}, \bibinfo {author} {\bibfnamefont {L.~E.}\ \bibnamefont {Silbert}},
  \ and\ \bibinfo {author} {\bibfnamefont {M.}~\bibnamefont {Sperl}},\ }\href
  {\doibase 10.1103/PhysRevLett.116.068302} {\bibfield  {journal} {\bibinfo
  {journal} {Phys. Rev. Lett.}\ }\textbf {\bibinfo {volume} {116}},\ \bibinfo
  {pages} {068302} (\bibinfo {year} {2016})}\BibitemShut {NoStop}%
\bibitem [{\citenamefont {Mizuno}()}]{Hideyuki}%
  \BibitemOpen
  \bibfield  {author} {\bibinfo {author} {\bibfnamefont {H.}~\bibnamefont
  {Mizuno}},\ }\href@noop {} {}\bibinfo {howpublished} {personal
  communication}\BibitemShut {NoStop}%
\bibitem [{\citenamefont {Chikkadi}\ \emph {et~al.}(2011)\citenamefont
  {Chikkadi}, \citenamefont {Wegdam}, \citenamefont {Bonn}, \citenamefont
  {Nienhuis},\ and\ \citenamefont {Schall}}]{chikkadi2011long}%
  \BibitemOpen
  \bibfield  {author} {\bibinfo {author} {\bibfnamefont {V.}~\bibnamefont
  {Chikkadi}}, \bibinfo {author} {\bibfnamefont {G.}~\bibnamefont {Wegdam}},
  \bibinfo {author} {\bibfnamefont {D.}~\bibnamefont {Bonn}}, \bibinfo {author}
  {\bibfnamefont {B.}~\bibnamefont {Nienhuis}}, \ and\ \bibinfo {author}
  {\bibfnamefont {P.}~\bibnamefont {Schall}},\ }\href@noop {} {\bibfield
  {journal} {\bibinfo  {journal} {Physical Review Letters}\ }\textbf {\bibinfo
  {volume} {107}},\ \bibinfo {pages} {198303} (\bibinfo {year}
  {2011})}\BibitemShut {NoStop}%
\bibitem [{\citenamefont {Shang}\ \emph {et~al.}(2014)\citenamefont {Shang},
  \citenamefont {Li}, \citenamefont {Yao}, \citenamefont {Lu},\ and\
  \citenamefont {Wang}}]{shang2014evolution}%
  \BibitemOpen
  \bibfield  {author} {\bibinfo {author} {\bibfnamefont {B.}~\bibnamefont
  {Shang}}, \bibinfo {author} {\bibfnamefont {M.}~\bibnamefont {Li}}, \bibinfo
  {author} {\bibfnamefont {Y.}~\bibnamefont {Yao}}, \bibinfo {author}
  {\bibfnamefont {Y.}~\bibnamefont {Lu}}, \ and\ \bibinfo {author}
  {\bibfnamefont {W.}~\bibnamefont {Wang}},\ }\href@noop {} {\bibfield
  {journal} {\bibinfo  {journal} {Physical Review E}\ }\textbf {\bibinfo
  {volume} {90}},\ \bibinfo {pages} {042303} (\bibinfo {year}
  {2014})}\BibitemShut {NoStop}%
\bibitem [{\citenamefont {Henkes}\ and\ \citenamefont
  {Chakraborty}(2009)}]{henkes_statistical_2009}%
  \BibitemOpen
  \bibfield  {author} {\bibinfo {author} {\bibfnamefont {S.}~\bibnamefont
  {Henkes}}\ and\ \bibinfo {author} {\bibfnamefont {B.}~\bibnamefont
  {Chakraborty}},\ }\href {\doibase 10.1103/PhysRevE.79.061301} {\bibfield
  {journal} {\bibinfo  {journal} {Phys. Rev. E}\ }\textbf {\bibinfo {volume}
  {79}},\ \bibinfo {pages} {061301} (\bibinfo {year} {2009})}\BibitemShut
  {NoStop}%
\bibitem [{\citenamefont {Goldhirsch}(2010)}]{goldhirsch2010stress}%
  \BibitemOpen
  \bibfield  {author} {\bibinfo {author} {\bibfnamefont {I.}~\bibnamefont
  {Goldhirsch}},\ }\href@noop {} {\bibfield  {journal} {\bibinfo  {journal}
  {Granular Matter}\ }\textbf {\bibinfo {volume} {12}},\ \bibinfo {pages} {239}
  (\bibinfo {year} {2010})}\BibitemShut {NoStop}%
\bibitem [{\citenamefont {Goldenberg}\ and\ \citenamefont
  {Goldhirsch}(2006)}]{goldenberg2006continuum}%
  \BibitemOpen
  \bibfield  {author} {\bibinfo {author} {\bibfnamefont {C.}~\bibnamefont
  {Goldenberg}}\ and\ \bibinfo {author} {\bibfnamefont {I.}~\bibnamefont
  {Goldhirsch}},\ }\href@noop {} {\  (\bibinfo {year} {2006})}\BibitemShut
  {NoStop}%
\bibitem [{\citenamefont {Plimpton}(1995)}]{Plimpton:1995:FPA:201627.201628}%
  \BibitemOpen
  \bibfield  {author} {\bibinfo {author} {\bibfnamefont {S.~J.}\ \bibnamefont
  {Plimpton}},\ }\href {\doibase 10.1006/jcph.1995.1039} {\bibfield  {journal}
  {\bibinfo  {journal} {J. Comput. Phys.}\ }\textbf {\bibinfo {volume} {117}},\
  \bibinfo {pages} {1} (\bibinfo {year} {1995})}\BibitemShut {NoStop}%
\bibitem [{\citenamefont {Silbert}\ \emph {et~al.}(2005)\citenamefont
  {Silbert}, \citenamefont {Liu},\ and\ \citenamefont
  {Nagel}}]{silbert_vibrations_2005}%
  \BibitemOpen
  \bibfield  {author} {\bibinfo {author} {\bibfnamefont {L.~E.}\ \bibnamefont
  {Silbert}}, \bibinfo {author} {\bibfnamefont {A.~J.}\ \bibnamefont {Liu}}, \
  and\ \bibinfo {author} {\bibfnamefont {S.~R.}\ \bibnamefont {Nagel}},\ }\href
  {\doibase 10.1103/PhysRevLett.95.098301} {\bibfield  {journal} {\bibinfo
  {journal} {Phys. Rev. Lett.}\ }\textbf {\bibinfo {volume} {95}},\ \bibinfo
  {pages} {098301} (\bibinfo {year} {2005})}\BibitemShut {NoStop}%
\bibitem [{\citenamefont {Goodrich}\ \emph {et~al.}(2014)\citenamefont
  {Goodrich}, \citenamefont {Liu},\ and\ \citenamefont
  {Nagel}}]{goodrich_solids_2014}%
  \BibitemOpen
  \bibfield  {author} {\bibinfo {author} {\bibfnamefont {C.~P.}\ \bibnamefont
  {Goodrich}}, \bibinfo {author} {\bibfnamefont {A.~J.}\ \bibnamefont {Liu}}, \
  and\ \bibinfo {author} {\bibfnamefont {S.~R.}\ \bibnamefont {Nagel}},\ }\href
  {\doibase 10.1038/nphys3006} {\bibfield  {journal} {\bibinfo  {journal} {Nat
  Phys}\ }\textbf {\bibinfo {volume} {10}},\ \bibinfo {pages} {578} (\bibinfo
  {year} {2014})}\BibitemShut {NoStop}%
\bibitem [{\citenamefont {O'Hern}\ \emph {et~al.}(2003)\citenamefont {O'Hern},
  \citenamefont {Silbert}, \citenamefont {Liu},\ and\ \citenamefont
  {Nagel}}]{o2003jamming}%
  \BibitemOpen
  \bibfield  {author} {\bibinfo {author} {\bibfnamefont {C.~S.}\ \bibnamefont
  {O'Hern}}, \bibinfo {author} {\bibfnamefont {L.~E.}\ \bibnamefont {Silbert}},
  \bibinfo {author} {\bibfnamefont {A.~J.}\ \bibnamefont {Liu}}, \ and\
  \bibinfo {author} {\bibfnamefont {S.~R.}\ \bibnamefont {Nagel}},\ }\href@noop
  {} {\bibfield  {journal} {\bibinfo  {journal} {Physical Review E}\ }\textbf
  {\bibinfo {volume} {68}},\ \bibinfo {pages} {011306} (\bibinfo {year}
  {2003})}\BibitemShut {NoStop}%
\bibitem [{\citenamefont {Lerner}\ \emph {et~al.}(2014)\citenamefont {Lerner},
  \citenamefont {DeGiuli}, \citenamefont {D{\"u}ring},\ and\ \citenamefont
  {Wyart}}]{lerner2014breakdown}%
  \BibitemOpen
  \bibfield  {author} {\bibinfo {author} {\bibfnamefont {E.}~\bibnamefont
  {Lerner}}, \bibinfo {author} {\bibfnamefont {E.}~\bibnamefont {DeGiuli}},
  \bibinfo {author} {\bibfnamefont {G.}~\bibnamefont {D{\"u}ring}}, \ and\
  \bibinfo {author} {\bibfnamefont {M.}~\bibnamefont {Wyart}},\ }\href@noop {}
  {\bibfield  {journal} {\bibinfo  {journal} {Soft matter}\ }\textbf {\bibinfo
  {volume} {10}},\ \bibinfo {pages} {5085} (\bibinfo {year}
  {2014})}\BibitemShut {NoStop}%
\bibitem [{\citenamefont {Yoshimoto}\ \emph {et~al.}(2004)\citenamefont
  {Yoshimoto}, \citenamefont {Jain}, \citenamefont {Workum}, \citenamefont
  {Nealey},\ and\ \citenamefont {de~Pablo}}]{yoshimoto_mechanical_2004}%
  \BibitemOpen
  \bibfield  {author} {\bibinfo {author} {\bibfnamefont {K.}~\bibnamefont
  {Yoshimoto}}, \bibinfo {author} {\bibfnamefont {T.~S.}\ \bibnamefont {Jain}},
  \bibinfo {author} {\bibfnamefont {K.~V.}\ \bibnamefont {Workum}}, \bibinfo
  {author} {\bibfnamefont {P.~F.}\ \bibnamefont {Nealey}}, \ and\ \bibinfo
  {author} {\bibfnamefont {J.~J.}\ \bibnamefont {de~Pablo}},\ }\href {\doibase
  10.1103/PhysRevLett.93.175501} {\bibfield  {journal} {\bibinfo  {journal}
  {Phys. Rev. Lett.}\ }\textbf {\bibinfo {volume} {93}},\ \bibinfo {pages}
  {175501} (\bibinfo {year} {2004})}\BibitemShut {NoStop}%
\bibitem [{\citenamefont {Mizuno}\ \emph {et~al.}(2013)\citenamefont {Mizuno},
  \citenamefont {Mossa},\ and\ \citenamefont {Barrat}}]{PhysRevE.87.042306}%
  \BibitemOpen
  \bibfield  {author} {\bibinfo {author} {\bibfnamefont {H.}~\bibnamefont
  {Mizuno}}, \bibinfo {author} {\bibfnamefont {S.}~\bibnamefont {Mossa}}, \
  and\ \bibinfo {author} {\bibfnamefont {J.-L.}\ \bibnamefont {Barrat}},\
  }\href {\doibase 10.1103/PhysRevE.87.042306} {\bibfield  {journal} {\bibinfo
  {journal} {Physical Review E}\ }\textbf {\bibinfo {volume} {87}},\ \bibinfo
  {pages} {042306} (\bibinfo {year} {2013})}\BibitemShut {NoStop}%
\bibitem [{\citenamefont {Tanaka}\ \emph {et~al.}(2010)\citenamefont {Tanaka},
  \citenamefont {Kawasaki}, \citenamefont {Shintani},\ and\ \citenamefont
  {Watanabe}}]{tanaka_critical-like_2010}%
  \BibitemOpen
  \bibfield  {author} {\bibinfo {author} {\bibfnamefont {H.}~\bibnamefont
  {Tanaka}}, \bibinfo {author} {\bibfnamefont {T.}~\bibnamefont {Kawasaki}},
  \bibinfo {author} {\bibfnamefont {H.}~\bibnamefont {Shintani}}, \ and\
  \bibinfo {author} {\bibfnamefont {K.}~\bibnamefont {Watanabe}},\ }\href
  {\doibase 10.1038/nmat2634} {\bibfield  {journal} {\bibinfo  {journal} {Nat
  Mater}\ }\textbf {\bibinfo {volume} {9}},\ \bibinfo {pages} {324} (\bibinfo
  {year} {2010})}\BibitemShut {NoStop}%
\end{thebibliography}
%
\end{document}